\documentclass[sigconf]{acmart}

\AtBeginDocument{%
  \providecommand\BibTeX{{%
    \normalfont B\kern-0.5em{\scshape i\kern-0.25em b}\kern-0.8em\TeX}}}

\setcopyright{acmcopyright}
\copyrightyear{2023}
\acmYear{2023}
\acmDOI{XXXXXXX.XXXXXXX}

\acmConference[Conference acronym 'XX]{Make sure to enter the correct
  conference title from your rights confirmation email}{June 03--05,
  2018}{Woodstock, NY}
\acmPrice{15.00}
\acmISBN{978-1-4503-XXXX-X/18/06}

\usepackage{tipa}
\usepackage{multirow}
\usepackage{array}
\usepackage{booktabs}
\usepackage{subcaption}
\usepackage{makecell}

\makeatletter                   
\def\mdseries@tt{m}             
\makeatother                    
\usepackage[plain]{fancyref}
\usepackage[draft=true]{minted} 
\usepackage{color}
\usepackage{hyperref}           
\hypersetup{
    colorlinks=true,
    linkcolor=blue,
    filecolor=red,      
    urlcolor=magenta,
    breaklinks=true,            
}
\usepackage{breakurl}           

\begin{document}
\sloppy                         

\title{IPA-CLIP: Integrating Phonetic Priors into Vision and Language Pretraining}

\author{Chihaya Matsuhira}
\email{matsuhirac@cs.is.i.nagoya-u.ac.jp}
\orcid{0000-0003-2453-4560}
\affiliation{%
  \institution{Nagoya University}
  \state{Aichi}
  \country{Japan}
}
\author{Marc A. Kastner}
\email{mkastner@i.kyoto-u.ac.jp}
\orcid{0000-0002-9193-5973}
\affiliation{%
  \institution{Kyoto University}
  \state{Kyoto}
  \country{Japan}
}
\author{Takahiro Komamizu}
\email{taka-coma@acm.org}
\orcid{0000-0002-3041-4330}
\affiliation{%
  \institution{Nagoya University}
  \state{Aichi}
  \country{Japan}
}
\author{Takatsugu Hirayama}
\email{t-hirayama@uhe.ac.jp}
\orcid{0000-0001-6290-9680}
\affiliation{%
  \institution{University of Human Environments}
  \state{Aichi}
  \country{Japan}
}
\author{Keisuke Doman}
\email{kdoman@sist.chukyo-u.ac.jp}
\orcid{0000-0001-6040-4988}
\affiliation{%
  \institution{Chukyo University}
  \state{Aichi}
  \country{Japan}
}
\author{Yasutomo Kawanishi}
\email{yasutomo.kawanishi@riken.jp}
\orcid{0000-0002-3799-4550}
\affiliation{%
  \institution{RIKEN}
  \state{Kyoto}
  \country{Japan}
}
\author{Ichiro Ide}
\email{ide@i.nagoya-u.ac.jp}
\orcid{0000-0003-3942-9296}
\affiliation{%
  \institution{Nagoya University}
  \state{Aichi}
  \country{Japan}
}

\renewcommand{\shortauthors}{C. Matsuhira et al.}

\begin{abstract}
Recently, large-scale Vision and Language (V\&L) pretraining has become the standard backbone of many multimedia systems.
While it has shown remarkable performance even in unseen situations, it often performs in ways not intuitive to humans.
Particularly, they usually do not consider the pronunciation of the input, which humans would utilize to understand language, especially when it comes to unknown words.
Thus, this paper inserts phonetic prior into Contrastive Language-Image Pretraining (CLIP), one of the V\&L pretrained models, to make it consider the pronunciation similarity among its pronunciation inputs.
To achieve this, we first propose a phoneme embedding that utilizes the phoneme relationships provided by the International Phonetic Alphabet (IPA) chart as a phonetic prior.
Next, by distilling the frozen CLIP text encoder, we train a pronunciation encoder employing the IPA-based embedding. The proposed model named IPA-CLIP comprises this pronunciation encoder and the original CLIP encoders (image and text).
Quantitative evaluation reveals that the phoneme distribution on the embedding space represents phonetic relationships more accurately when using the proposed phoneme embedding.
Furthermore, in some multimodal retrieval tasks, we confirm that the proposed pronunciation encoder enhances the performance of the text encoder and that the pronunciation encoder handles nonsense words in a more phonetic manner than the text encoder.
Finally, qualitative evaluation verifies the correlation between the pronunciation encoder and human perception regarding pronunciation similarity.
\end{abstract}

\ccsdesc[500]{Information systems~Language models}
\ccsdesc[100]{Information systems~Similarity measures}
\ccsdesc[300]{Computing methodologies~Image representations}
\ccsdesc[500]{Computing methodologies~Phonology / morphology}
\ccsdesc[300]{Applied computing~Psychology}
\keywords{pronunciation, phonetics, phonology, vision and language, distillation, multimodal understanding}

\maketitle

\begin{figure*}[t]
  \begin{tabular}{cc}
    \begin{minipage}[t]{0.6\hsize}
      \centering
      \includegraphics[width = 1\columnwidth]{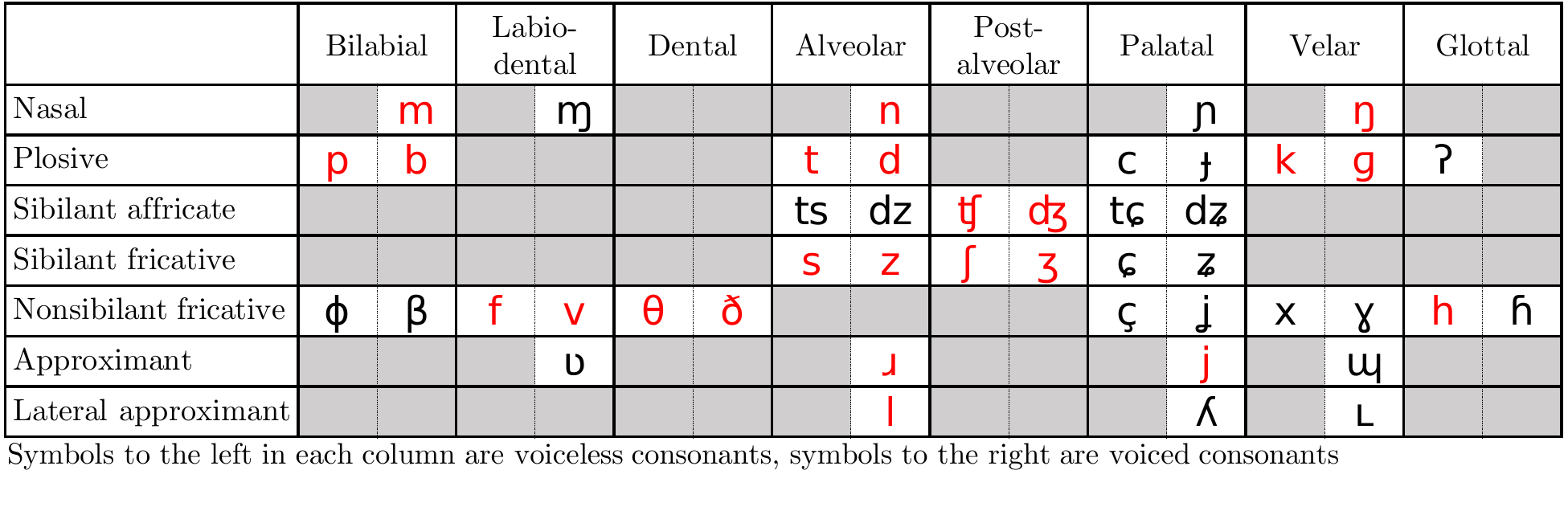}
      \subcaption{Consonants (Pulmonic)}
      \label{fig:ipachart:cons}
    \end{minipage} &
    \hspace{0.1\columnwidth}
    \begin{minipage}[t]{0.3\hsize}
      \centering
      \includegraphics[width = 1\columnwidth]{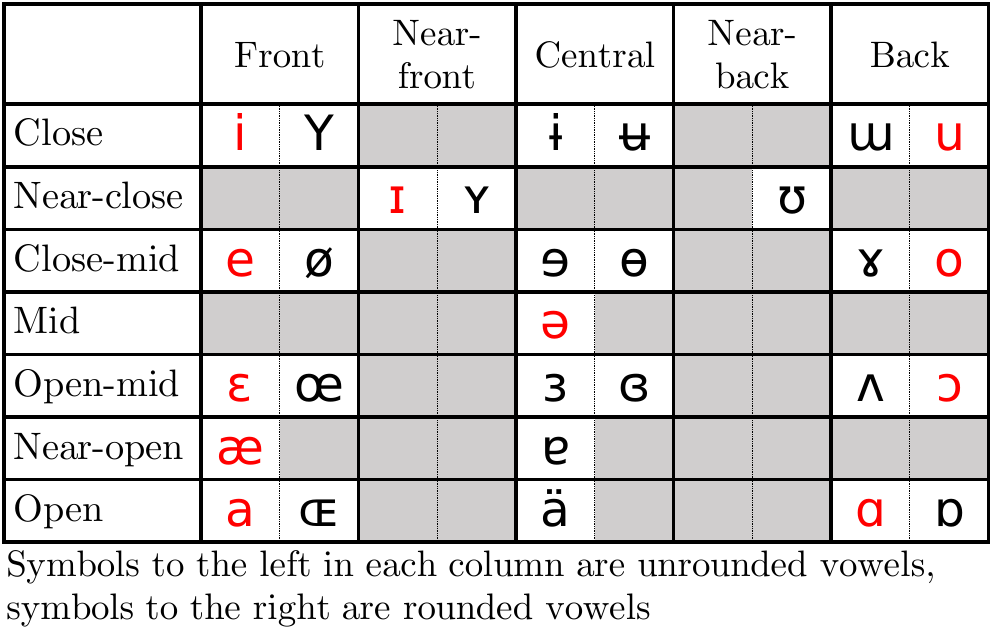}
      \subcaption{Vowels}
      \label{fig:ipachart:vows}
    \end{minipage}
  \end{tabular}
  \caption[]{IPA Chart~\cite{bib:ipachart} for pulmonic consonants and vowels used in this paper. It connects almost all phonemes occurring in natural languages regarding their phonetic relationships. English phonemes, as used in this paper, are colored red.}
  \label{fig:ipachart}
\end{figure*}

\section{Introduction}\label{sec:intro}

Vision and Language (V\&L) pretraining from large-scale image-text datasets has gained increasing attention as a fundamental model of multimedia systems. Contrastive Language-Image Pretraining (CLIP)~\cite{bib:clip} is one of such V\&L pretrained models consisting of an image encoder and a text encoder that share their bi-modal embedding space.
It uses an order-of-magnitude larger training data than previous models, which guarantees its effectiveness in various applications including image classification and retrieval~\cite{bib:clip}, object detection~\cite{bib:proposalclip}, image generation~\cite{bib:vqganclip}, and image captioning~\cite{bib:clipglass}. 
This also allows it to perform well even in scenarios not seen in the training set.

However, in many cases, such models do not behave in a way intuitive to humans. 
One of the reasons is that they do not consider the phonetic features of the inputs, i.e. how words are pronounced, which humans would consciously or unconsciously utilize to intuitively express the meanings of words. 
For example, when an English speaker uses the word ``\textit{Lump}'' in a conversation, they might have a connotation of something \textbf{heavy and round}, akin to other similar-sounding words ``\textit{Bump}'', ``\textit{Slump}'', and ``\textit{Plump}''.

Humans also use such phonetic knowledge to process spoken language~\cite{bib:hahnbailey}, especially when they hear unknown words or nonsense words (in short, nonwords).
Humans tend to imagine the meanings of such nonwords by associating them with their similar-sounding words.
For instance, when English speakers hear the pronunciation of a nonword ``\textit{Britch}'', they might associate it with the word ``\textit{Bridge}'', thus the meaning of ``\textit{Britch}'' might be recognized as something related to a bridge.
Meanwhile, another nonword ``\textit{Brish}'' (rhymes with ``\textit{Fish}'') might be less perceived so because of its less phonetic similarity to ``\textit{Bridge}''. 
Without knowing phonetic relationships, conventional models and systems could not consider such correspondences.

\begin{figure*}[t]
  \begin{center}
      \includegraphics[width = 0.92\textwidth]{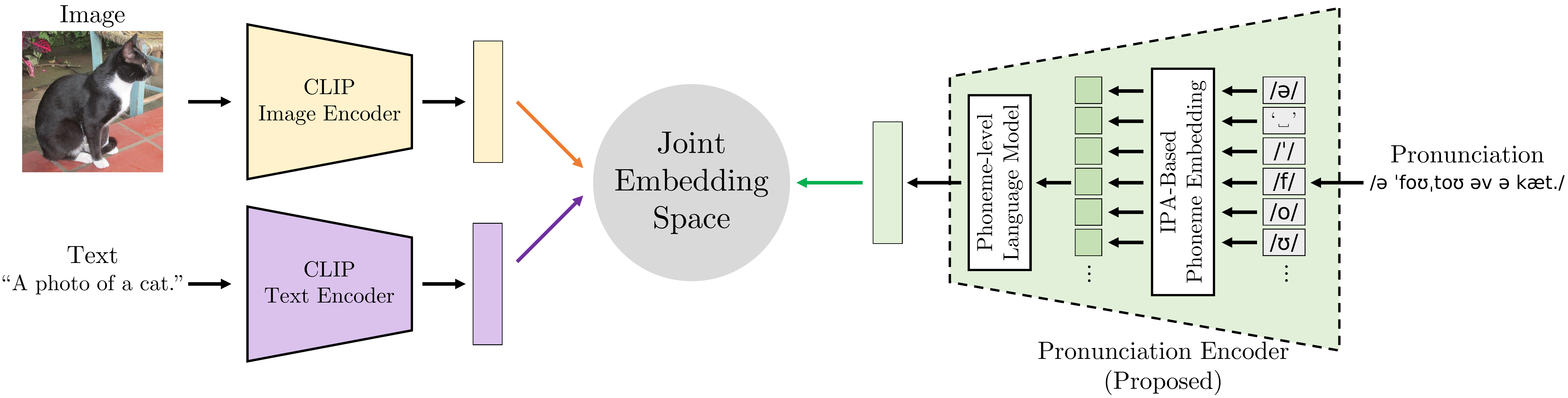}
  \end{center}
  \caption[]{Overview of the proposed IPA-CLIP model.}
  \label{fig:flow}
\end{figure*}

The goal of this paper is thus to insert phonetic priors into V\&L pretrained models to make them consider the phonetic similarity among pronunciation inputs. By doing this, we also aim to enable the models to handle inputs containing nonwords whose pronunciations are slightly different from certain existing words. This will make them better correspond to human expectations towards such inputs. 

A possible approach to insert phonetic knowledge into the pretrained models is to change the tokenizer of the text encoders and retrain or finetune the whole model with image-text pairs.
In the CLIP model, for example, Byte Pair Encoding~\cite{bib:bpe} is employed to tokenize and embed text inputs. However, this does not handle nonwords in the aforementioned way. 
Other existing embeddings, such as character-level embedding~\cite{bib:characterbert,bib:charbert}, which is more robust mainly against spelling mistakes, could be more effective.
Yet, these types of embeddings are still not sufficient since they also ignore phonetic similarity. One obstacle to this is that their language input is written with graphemes, which do not necessarily correspond to phonemes. 
Furthermore, retraining or finetuning the original pretrained model could be another drawback since it requires both a number of image-text pairs and a huge computational cost.

To tackle these problems, we first integrate a phonetic prior into a general phoneme-level embedding.
We utilize the International Phonetic Alphabet (IPA) chart~\cite{bib:ipachart}, as shown in Fig.~\ref{fig:ipachart}, to propose an IPA-based phoneme embedding. The chart connects almost all phonemes that can appear in any natural language and the proximity on it indicates phonetic similarities.
The proposed method incorporates these relationships on the IPA chart for the creation of an embedding space. 
While the phonetic prior and thus the proposed phoneme embedding should work for any language, in this paper, the rest part of the proposed method is primarily designed, implemented, and evaluated for the English language.

Next, we take a distillation approach to extend a V\&L pretrained model to accept language inputs written with phonemes.
Specifically, we implement IPA-CLIP as illustrated in Fig.~\ref{fig:flow}, a model which extends CLIP. IPA-CLIP consists of three encoders: the original CLIP image and text encoders, and a newly trained pronunciation encoder.
The pronunciation encoder takes an array of phonemes written with IPA symbols as an input (e.g., \textipa{/@ "foU""toU @v @ k\ae t./} for ``A photo of a cat.''). This allows the IPA-based phoneme embedding in the encoder to process each phoneme based on its phonetic knowledge.
To train this encoder, we freeze and distill the original CLIP text encoder using a number of text-pronunciation pairs.
This distillation reduces the cost of obtaining the pronunciation encoder while also sharing the feature space with the bi-modal joint embedding space of the original CLIP. 
Moreover, sharing the embedding space can widen the application range of CLIP. 
Since the pronunciation encoder is trained via distillation, any system relying on the CLIP text encoder will be able to accept also pronunciation inputs of languages even if they lack orthography, just by substituting the two encoders.

The contributions of this paper can be summarized as follows: (1)~We propose an IPA-based phoneme embedding which integrates phonetic similarity on the IPA chart into its phoneme embedding space, (2)~We implement IPA-CLIP, a model which extends CLIP~\cite{bib:clip}, one of the vision and language pretrained models, to accept an arbitrary pronunciation input by utilizing the IPA-based phoneme embedding, (3)~We confirm the agreement of the outputs of the IPA-based phoneme embedding with the phonetic relationships on the IPA chart, (4)~We apply IPA-CLIP to some retrieval tasks to discuss its performance compared with the original CLIP and its ability to relate unknown words with their similar-sounding known words, and (5)~We evaluate the correlation between the embedding space of IPA-CLIP and human perception regarding pronunciation similarity.

In Section~\ref{sec:phonetics}, we first introduce the phonetic relationships provided by phonetics, before referring to related studies in Section~\ref{sec:related}.
Section~\ref{sec:method} describes the details of the proposed IPA-based phoneme embedding and IPA-CLIP. We report experiments and discuss the results in Section~\ref{sec:exp} and Section~\ref{sec:exp2}, followed by the conclusion of this paper in Section~\ref{sec:conclusion}.

\section{Phoneme Attributes in Phonetics}\label{sec:phonetics}

Alphabets to describe the pronunciation of words are called phonetic symbols.
IPA (International Phonetic Alphabet) transcription is one of the most common alphabetic systems used to describe pronunciation. It assigns a unique symbol to each phoneme while providing symbols for phonetic components other than phonemes, such as stresses and syllable boundaries.
The IPA chart~\cite{bib:ipachart}, as shown in Fig.~\ref{fig:ipachart}, lists all IPA symbols and denotes the relationships among all phonemes that can appear in any natural language. 
On the chart, each phoneme is characterized by multiple phonetic attributes. Consonants have three attributes: \textit{voicing}, \textit{place of articulation}, and \textit{manner of articulation}. Vowels also have three attributes: \textit{height}, \textit{backness}, and \textit{roundedness}. According to the chart, for example, the voiceless velar plosive \textipa{/k/}, as in ``\textit{\textbf{C}oat}'', possesses ``voiceless'' (\textit{voicing}), ``velar'' (\textit{place}), and ``plosive'' (\textit{manner}) consonant attributes, and the close-mid back rounded vowel \textipa{/o/}, as in ``\textit{C\textbf{o}at}'', possesses ``close-mid'' (\textit{height}), ``back'' (\textit{backness}), and ``rounded'' (\textit{roundedness}) vowel attributes. Some consonants such as the voiced labial–velar approximant \textipa{/w/} have multiple places of articulation. In this case, \textipa{/w/} is characterized by four consonant attributes (``voiced'', ``labial'', ``velar'', and ``approximant'').

This paper incorporates these phonetic attributes to construct the embedding space of the proposed IPA-based phoneme embedding. 

\section{Related Work}\label{sec:related}

Section~\ref{sec:related:computational} introduces several pieces of previous work that have utilized such phonetic information to construct phonetic spaces by means of computational approaches. Next, Section~\ref{sec:related:clipextension} introduces several methods which attempt to extend CLIP for other modalities.

\subsection{Computational Approaches to Phonetics}\label{sec:related:computational}

Several existing studies integrate phonetic knowledge, such as the IPA chart, into the calculation of the phonetic similarity between words written with letters or phonetic symbols~\cite{bib:vitzwinkler,bib:hahnbailey,bib:parrish,bib:bay}. 
Vitz and Winkler~\cite{bib:vitzwinkler} propose \textit{Predicted Phonemic Distance}, which is a dissimilarity between the pronunciations of two words calculated based on a method similar to the edit distance. Although it is directed at predicting the dissimilarity among pronunciation, it does not consider the phonetic relationships. 
To tackle this, Hahn and Bailey~\cite{bib:hahnbailey} incorporate phonetic features of phonemes into the edit distance measure. Their method regards the distance of two phonemes sharing certain attributes (e.g. the voiceless \textbf{velar} \textbf{plosive} \textipa{/k/} and the voiced \textbf{velar} \textbf{plosive} \textipa{/g/}) as closer than other pairs of phonemes sharing fewer attributes (e.g. the voiceless velar plosive \textipa{/k/} and the voiced bilabial nasal \textipa{/m/}). They compare the distance measure with human perception and show its effectiveness, especially for monosyllabic words. 
Parrish~\cite{bib:parrish} proposes a bi-gram model for poetic applications.
The model is constructed based on phonetic features similar to the ones listed on the IPA chart. The author extracts 949 unique bi-grams of phonetic features that can occur within English words and make histograms based on them. The phonetic similarity is measured as the cosine similarity between histograms for words.
Bay et al.~\cite{bib:bay} make use of the structure of the IPA chart to calculate the phonetic similarity, which they call the rhyme score, for text transformation. 
They regard all three consonant attributes, as described in Section~\ref{sec:phonetics}, as categorical attributes, and \textit{height} and \textit{backness} of the vowel attributes as continuous attributes. 
Particularly, to calculate the similarity between consonants, they check which attributes two consonants have in common and sum the Kronecker deltas of the commonness weighted with empirical weights. For vowels, they manually reconstruct the vowel chart on a 2D Cartesian plane (they ignore \textit{roundedness}) and measure the Euclidean distance between vowels. 

Recent Natural Language Processing (NLP) techniques also obtain neural phoneme embeddings that reflect phonetic similarity without explicit prior and supervision~\cite{bib:kolachinamagyar,bib:boldsen}.
Kolachina and Magyar~\cite{bib:kolachinamagyar} evaluate if Word2vec~\cite{bib:word2vec1,bib:word2vec2} can learn the phonetic relationships among phonemes. They feed the model with corpora of artificial languages as well as English, concluding that Word2vec embeddings capture the phonetic relationships from such corpora quite well. Boldsen et al.~\cite{bib:boldsen} perform a similar analysis of character embeddings in multiple languages using Long Short-Term Memory (LSTM) and Transformer by training these models for the next character prediction task. The obtained character embeddings show strong correlations to the sound representations derived from phoneme classifications in English, which supports the results of the previous work~\cite{bib:kolachinamagyar}.

For constructing the IPA-based phoneme embedding in this paper, we follow the usage of the IPA chart as proposed in Bay et al.~\cite{bib:bay} and treat two vowel attributes as continuous while other attributes as categorical. We also compare the IPA-based embedding with a neural phoneme embedding obtained via training without such a prior as a baseline.

The main difference between our goal and theirs is that our phonetic space considers not only phonetic but also semantic similarity. Thus, for instance, two words ``\textit{Bit}'' and ``\textit{Pit}'', which have totally different meanings, do not necessarily become neighbors on our embedding space, while they do in the methods introduced above.

\subsection{CLIP Extensions for Other Types of Data}\label{sec:related:clipextension}

Many methods are proposed to extend CLIP to other modalities because of its effectiveness for various multimodal tasks.
For the audio modality, several studies~\cite{bib:audioclip,bib:wav2clip,bib:clap} train a new audio encoder in addition to the original image and text encoders using multimodal datasets. 
Guzhov et al.~\cite{bib:audioclip} train three encoders for each modality simultaneously by utilizing multiple single and multimodal datasets.
Wu et al.~\cite{bib:wav2clip} freeze and distill the CLIP image encoder to train only an additional audio encoder with an audio-visual dataset. This enables the audio encoder to be compatible not only with the image encoder but also with the text encoder.
Elizalde et al.~\cite{bib:clap} use a text-audio dataset to train audio and text encoders from scratch.
All of these methods employ contrastive learning for training, which is derived from the original CLIP~\cite{bib:clip}.

Within the image and text modalities, Carlsson et al.~\cite{bib:multilingualclip} expand the CLIP text encoder, which was trained mainly on the English vocabulary, to be able to deal with multiple languages.
They first prepare a set of English sentences and then machine-translate them into multiple languages to obtain a multilingual dataset. Using the dataset, similar to Wu et al.~\cite{bib:wav2clip}, they train another multilingual text encoder with multilingual sentence pairs by freezing and distilling the CLIP text encoder. This guarantees that the outputs for the machine-translated sentences become similar to the ones for the original English sentences. During the training, instead of contrastive loss, they optimize Mean Squared Error (MSE) loss for the distillation of the CLIP text encoder.

In this paper, we extend CLIP for pronunciation inputs to propose IPA-CLIP.
To this end, we take a similar distillation approach as proposed in Carlsson et al.~\cite{bib:multilingualclip} to reduce the costs of training a pronunciation encoder compatible with the CLIP image encoder.
We automatically convert English sentences into phonetic transcriptions and then use the text-pronunciation pairs in place of the pairs of multilingual texts.

\section{IPA-CLIP: Phonetic Embedding Distillation of CLIP}\label{sec:method}

This paper proposes IPA-CLIP, an extension of CLIP for pronunciation inputs.
The overview of IPA-CLIP is illustrated in Fig.~\ref{fig:flow}. It consists of three encoders: The CLIP image encoder, the CLIP text encoder, and a new pronunciation encoder, all of which share the same multimodal embedding space.

In Section~\ref{sec:method:ipa}, we first introduce the method to embed the phonetic prior on the IPA chart into the phoneme embedding space.
Section~\ref{sec:method:distill} then explains the distillation process of IPA-CLIP to train a new pronunciation encoder employing the IPA-based embedding, followed by Section~\ref{sec:method:implement} which describes the implementation details of the proposed methods.

\begin{figure*}[t]
  \begin{tabular}{cc}
    \begin{minipage}[t]{0.51\hsize}
    \begin{center}
      \includegraphics[width = 1\columnwidth]{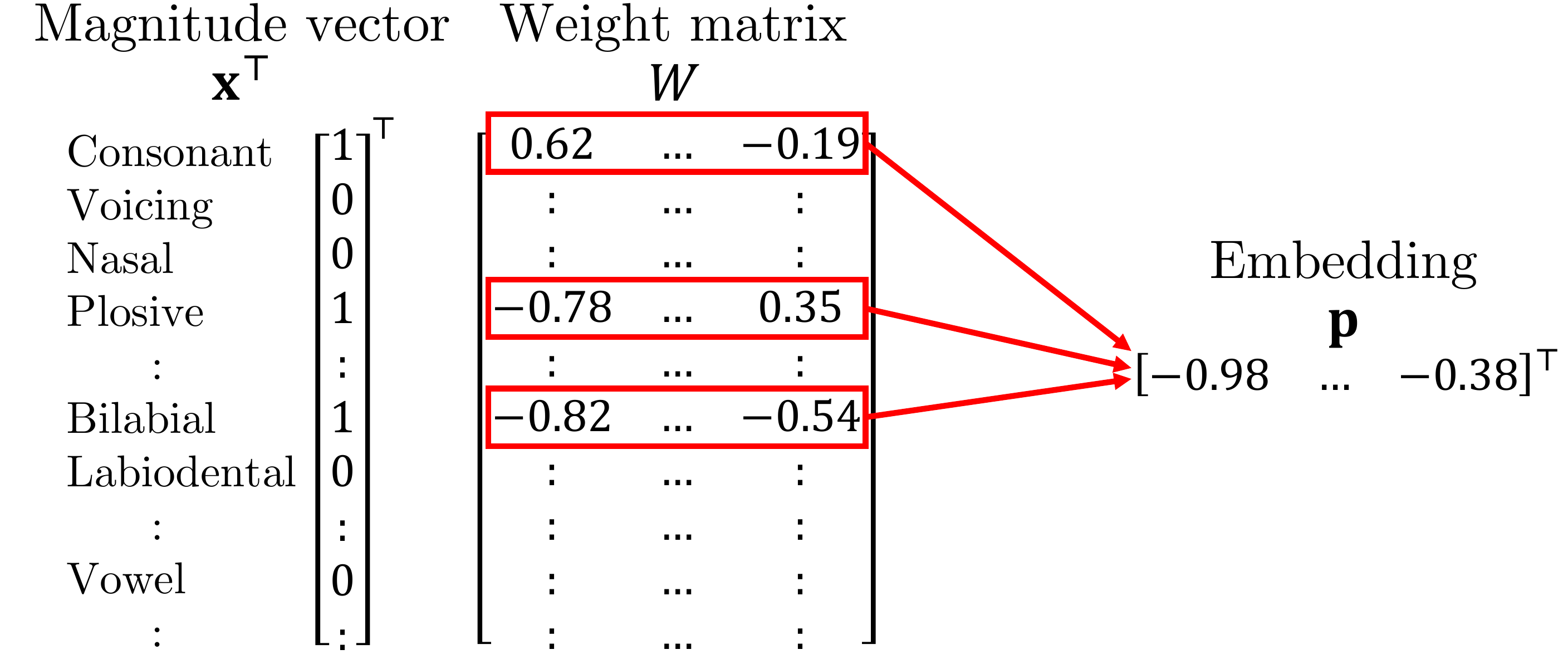}
      \subcaption{Proposed IPA-based phoneme embedding. The figure shows the calculation of the phoneme embedding for the voiceless bilabial plosive \textipa{/p/}.}
      \label{fig:method:ipaemb}
    \end{center}
    \end{minipage}
    \hspace{0.1\columnwidth}
    \begin{minipage}[t]{0.44\hsize}
    \begin{center}
      \includegraphics[width = 1\columnwidth]{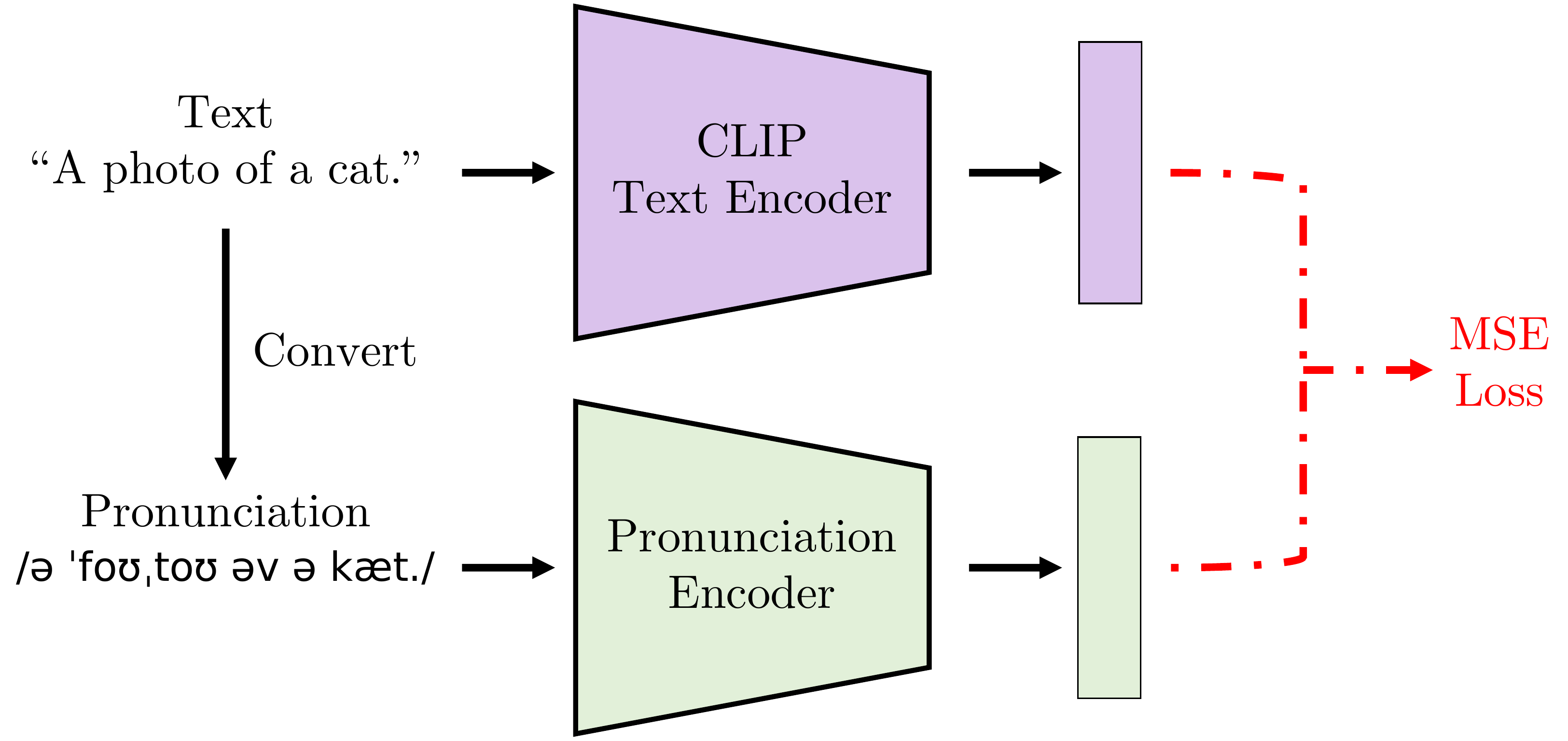}
      \subcaption{Distillation procedure.}
      \label{fig:method:distil}
    \end{center}
    \end{minipage}
  \end{tabular}
  \caption[]{Detailed illustration of the construction of the pronunciation encoder of IPA-CLIP.}
\end{figure*}

\subsection{IPA-based Phoneme Embedding}\label{sec:method:ipa}

We propose a phoneme embedding that considers the phonetic relationships among phonemes represented on the IPA chart. This phoneme embedding layer works by replacing the word embedding layer of any language model including BERT.
Because it is based on the IPA chart, the pronunciation input to this layer is theoretically universal and not specific to any language.

As mentioned in Section~\ref{sec:phonetics}, the IPA chart assigns three attributes for each phoneme: \textit{voicing}, \textit{place of articulation}, and \textit{manner of articulation} for consonants, and \textit{height}, \textit{backness}, and \textit{roundedness} for vowels.
Inspired by the literature utilizing these attributes~\cite{bib:bay}, we treat the two vowel attributes, \textit{height} and \textit{backness}, as continuous attributes. We consider the extent of the difference between these attributes among phonemes. For instance, the \textbf{close} front unrounded vowel \textipa{/i/} should be treated as more similar to the \textbf{close-mid} front unrounded vowel \textipa{/e/} than the \textbf{open} front unrounded vowel \textipa{/a/}. In contrast, following Bay et al.~\cite{bib:bay}, we regard the other four attributes as categorical attributes. We ignore the order of phonemes on the IPA chart and only focus on whether two phonemes have any attributes in common. The difference between the voiced bilabial nasal \textipa{/m/} and the voiceless velar plosive \textipa{/k/} is treated as equal to \textipa{/m/} and the voiceless alveolar plosives \textipa{/t/} although \textipa{/t/} is located closer to \textipa{/m/} on the chart.

As shown in Fig.~\ref{fig:method:ipaemb}, the proposed method calculates the phoneme embedding $\mathbf{p}$ as a linear combination $\sum_{i}{x_{i}\mathbf{w}_{i}}$, where $x_{i}$ is a magnitude and $\mathbf{w}_{i}$ is a feature vector for the $i$-th attribute.
In detail, for each phoneme, we calculate the multiplication of the transpose of the $N$-dimensional sparse magnitude vector $\mathbf{x}$ and the $N \times D$ feature matrix $W$, written as
\begin{equation}
    \mathbf{p} = \mathbf{x}^{\top}W = \sum_{i=1}^{N}{x_{i}\mathbf{w}_{i}} = x_{1}\mathbf{w}_{1} + x_{2}\mathbf{w}_{2} + \dots + x_{N}\mathbf{w}_{N}.
\end{equation}

Table~\ref{tab:listofattributes} shows examples of the $N$ attributes and magnitudes in the vector $\mathbf{x}$ for some phonemes. As shown in the table, $\mathbf{x}$ also includes attributes for letters other than phonemes such as stresses, spaces, commas, and exclamation marks. In the proposed method, we also project these letters onto the same phoneme embedding space although they are not phonemes.
The aim of this is to ensure the equivalent flexibility of the input of the pronunciation encoder to the CLIP text encoder. Thus, the pronunciation encoder can differentiate between homophonic texts such as ``everyday'' vs. ``every day'' and ``a cat'' vs. ``a cat!''.

This IPA-based phoneme embedding is used in the phoneme embedding layer of the pronunciation encoder explained in Section~\ref{sec:method:distill} to process pronunciation inputs.

\begin{table}[t]
\caption{Examples of attributes that each of the dimensions of the magnitude vector $\mathbf{x}$ represents.}\label{tab:listofattributes}
\begin{center}
\scalebox{0.8}{
\begin{tabular}{l|l|l|c|c|c|c|c}
\Xhline{0.08em}
\multicolumn{1}{c|}{\multirow{2}{*}{Attribute}} & \multicolumn{1}{c|}{\multirow{2}{*}{Category}} & \multicolumn{1}{c|}{\multirow{2}{*}{Range}} & \multicolumn{5}{c}{Examples of $\mathbf{x}$} \\ \cline{4-8}
& & & \textipa{/p/} & \textipa{/v/} & \textipa{/e/} & \textipa{/U/} & \textipa{`,'} \\
\hline
Consonant & \multirow{9}{*}{Consonant} & $x_{i} \in \{0,1\}$ & 1 & 1 & 0 & 0 & 0 \\ 
Voicing & & $x_{i} \in \{0,1\}$ & 0 & 1 & 0 & 0 & 0 \\
Manner 1: Nasal & & $x_{i} \in \{0,1\}$ & 0 & 0 & 0 & 0 & 0 \\
Manner 2: Plosive & & $x_{i} \in \{0,1\}$ & 1 & 0 & 0 & 0 & 0 \\
\multicolumn{1}{c|}{\resizebox*{!}{10pt}{$\vdots$}} &  & \multicolumn{1}{c|}{\resizebox*{!}{10pt}{$\vdots$}} & \multicolumn{1}{c|}{\resizebox*{!}{10pt}{$\vdots$}} & \multicolumn{1}{c|}{\resizebox*{!}{10pt}{$\vdots$}} & \multicolumn{1}{c|}{\resizebox*{!}{10pt}{$\vdots$}} & \multicolumn{1}{c|}{\resizebox*{!}{10pt}{$\vdots$}} & \multicolumn{1}{c}{\resizebox*{!}{10pt}{$\vdots$}} \\
Place 1: Bilabial & & $x_{i} \in \{0,1\}$ & 1 & 0 & 0 & 0 & 0 \\
Place 2: Labiodental & & $x_{i} \in \{0,1\}$ & 0 & 1 & 0 & 0 & 0 \\ 
\multicolumn{1}{c|}{\resizebox*{!}{10pt}{$\vdots$}} & & \multicolumn{1}{c|}{\resizebox*{!}{10pt}{$\vdots$}} & \multicolumn{1}{c|}{\resizebox*{!}{10pt}{$\vdots$}} & \multicolumn{1}{c|}{\resizebox*{!}{10pt}{$\vdots$}} & \multicolumn{1}{c|}{\resizebox*{!}{10pt}{$\vdots$}} & \multicolumn{1}{c|}{\resizebox*{!}{10pt}{$\vdots$}} & \multicolumn{1}{c}{\resizebox*{!}{10pt}{$\vdots$}} \\ \hline
Vowel & \multirow{4}{*}{Vowel} & $x_{i} \in \{0,1\}$ & 0 & 0 & 1 & 1 & 0 \\
Height & & $ 0 \leq x_{i} \leq 1$ & 0 & 0 & $\frac{2}{6}$ & $\frac{1}{6}$ & 0 \\
Backness & & $0 \leq x_{i} \leq 1$ & 0 & 0 & 0 & $\frac{3}{4}$ & 0 \\
Roundedness & & $x_{i} \in \{0,1\}$ & 0 & 0 & 0 & 1 & 0 \\ \hline
Primary stress \textipa{/"/} & \multirow{6}{*}{Others} & $x_{i} \in \{0,1\}$ & 0 & 0 & 0 & 0 & 0 \\
Secondary stress \textipa{/""/} &  & $x_{i} \in \{0,1\}$ & 0 & 0 & 0 & 0 & 0 \\
Char ` ': Space &  & $x_{i} \in \{0,1\}$ & 0 & 0 & 0 & 0 & 0 \\
Char `,': Comma &  & $x_{i} \in \{0,1\}$ & 0 & 0 & 0 & 0 & 1 \\ 
Char `!': Exclamation &  & $x_{i} \in \{0,1\}$ & 0 & 0 & 0 & 0 & 0 \\ 
\multicolumn{1}{c|}{\resizebox*{!}{10pt}{$\vdots$}} & & \multicolumn{1}{c|}{\resizebox*{!}{10pt}{$\vdots$}} & \multicolumn{1}{c|}{\resizebox*{!}{10pt}{$\vdots$}} & \multicolumn{1}{c|}{\resizebox*{!}{10pt}{$\vdots$}} & \multicolumn{1}{c|}{\resizebox*{!}{10pt}{$\vdots$}} & \multicolumn{1}{c|}{\resizebox*{!}{10pt}{$\vdots$}} & \multicolumn{1}{c}{\resizebox*{!}{10pt}{$\vdots$}} \\ \Xhline{0.08em}
\end{tabular}
}
\end{center}
\end{table}

\subsection{Training by Distilling CLIP Text Encoder}\label{sec:method:distill}

The training of the pronunciation encoder of IPA-CLIP is based on the distillation methods proposed by Carlsson et al.~\cite{bib:multilingualclip} and Wu et al.~\cite{bib:wav2clip}. Although the implementation of this paper focuses only on the English language, the distillation itself can be applied to other languages if resources are available.

The distillation procedure is illustrated in Fig.~\ref{fig:method:distil}.
First, to prepare a dataset of sentence-pronunciation pairs, we convert each sentence to its pronunciation with a given word-to-pronunciation dictionary.
Specifically, by looking up the dictionary, we replace all words in a sentence with the pronunciation of the words. In this conversion, for example, a sentence ``a photo of a cat.'' is converted to its pronunciation \textipa{/@ "foU""toU @v @ k\ae t./}. 
We ignore cases in the Latin alphabet and do not exclude letters other than the Latin alphabet as explained in Section~\ref{sec:method:ipa}.

With this dataset, we train our pronunciation encoder along with the CLIP text encoder.
The weights of the text encoder are frozen during the training. Given a sentence-pronunciation pair, we train the pronunciation encoder so that it can output the identical pronunciation embedding as the sentence embedding calculated by the text encoder. MSE loss is employed for the training objective as opposed to the contrastive loss used in training CLIP~\cite{bib:clip}. We choose the MSE loss because it is known to work better for the distillation purpose than the contrastive loss~\cite{bib:multilingualclip}. Moreover, training with the contrastive loss, i.e., the cosine similarity, yields embeddings similar in terms of the angle, but not in terms of the magnitude. Since several existing systems could require the exact magnitude of the CLIP embeddings, such a training objective could shrink the range of its application.

\subsection{Implementation}\label{sec:method:implement}

\begin{figure}[t]
  \begin{center}
      \includegraphics[width = 0.95\columnwidth]{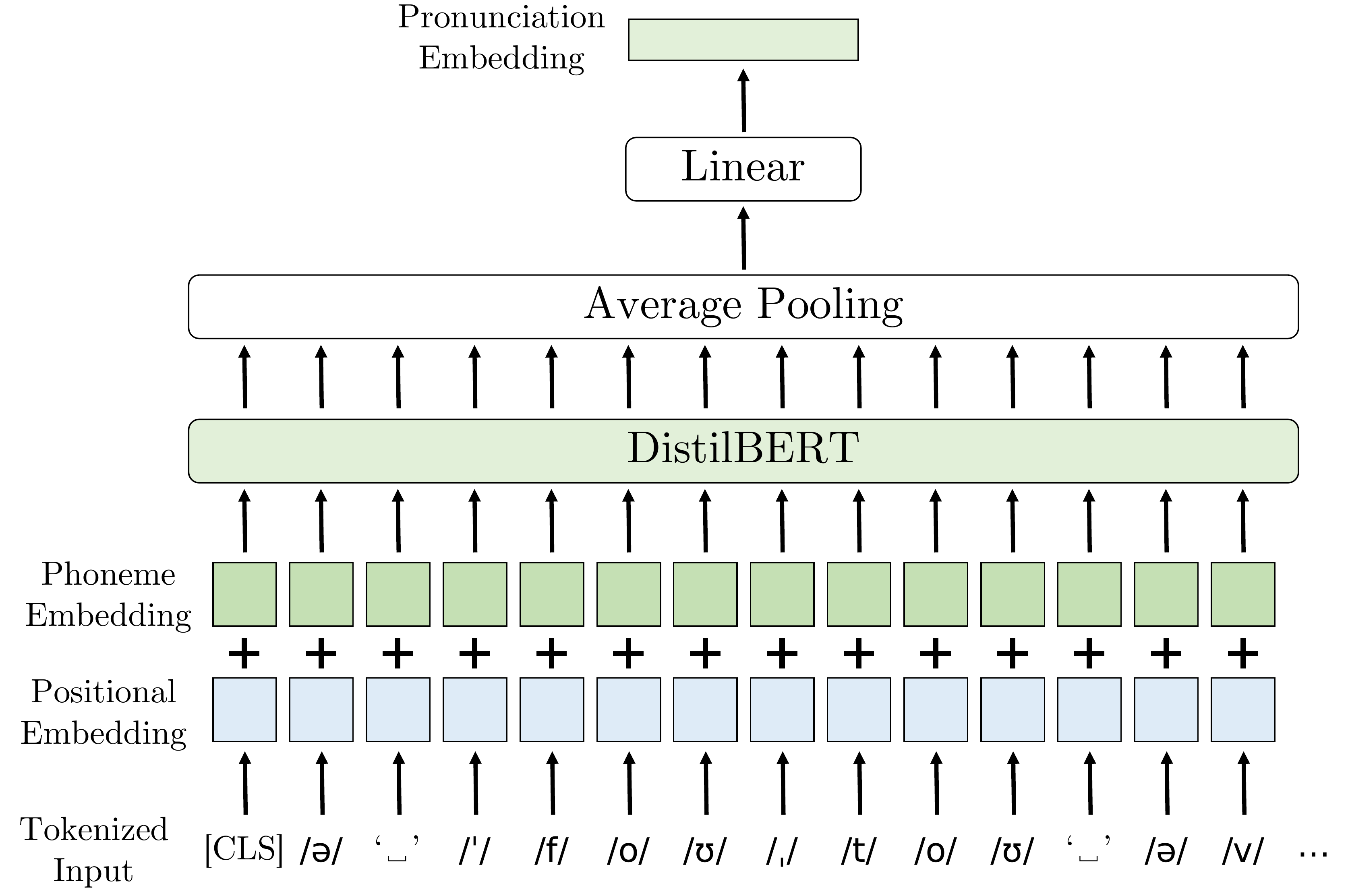}
  \end{center}
  \caption[]{Illustration of the pronunciation encoder used in IPA-CLIP. IPA-CLIP employs the proposed IPA-based phoneme embedding in its phoneme embedding layer.}
  \label{fig:implement}
\end{figure}

As the architecture of the pronunciation encoder, we adopt DistilBERT~\cite{bib:distilbert}, a light and efficient version of BERT. Figure~\ref{fig:implement} illustrates the implementation. 
We replace its word embedding with the proposed IPA-based phoneme embedding and add an additional linear layer to match the dimensionality of its output to that of the CLIP encoders, but we do not modify any other part of DistilBERT. The pronunciation encoder is trained from scratch.

To distill the original CLIP models that are trained primarily on English data, we use a list of English sentences compiled by Carlsson et al.~\cite{bib:multilingualclip}. It is a mixture of sentences taken from several image caption datasets, which could be strongly linked with the visual domain. In addition, to increase the vocabulary, we prepare sentences consisting of only one word by using Spell Checker Oriented Word Lists (SCOWL)\footnote{\url{http://wordlist.aspell.net/} (Accessed Jan. 19, 2023)}, an English wordlist that comprises 102,305 words. 
We convert these sentences into pronunciation written with IPA using the Python package eng-to-ipa\footnote{\url{https://pypi.org/project/eng-to-ipa/} (Accessed Jan. 19, 2023)}. The package uses the Carnegie-Mellon University (CMU) Pronouncing Dictionary, which is also used by many pieces of previous work~\cite{bib:parrish,bib:bay,bib:kolachinamagyar}. 
We remove sentences containing words whose pronunciations are not provided in the package. 
This results in training data of 1,168,451 sentences in total. Following the implementation of the previous work~\cite{bib:multilingualclip}, we fix the size of the validation split as 1,000, resulting in a split of 1,167,451 sentences for training and 1,000 sentences for validation.

We tested on two different pretrained CLIP models, ViT-B/32 and ViT-L/14, listed on the model card provided by OpenAI~\footnote{\url{https://github.com/openai/CLIP/blob/main/model-card.md} (Accessed Jan. 19, 2023)}. ViT-B/32 is the simplest and lightest model, while ViT-L/14 is the most recently released and most powerful model on the model card. These models employ Transformers for both image and text encoders.
We train our pronunciation encoder with a learning rate $5 \times 10^{-5}$, a batch size $32$, and the Adam optimizer~\cite{bib:adam}, up to $50$ epochs.

\section{Quantitative Evaluations}\label{sec:exp}

This section evaluates both the proposed IPA-based phoneme embedding and IPA-CLIP in a quantitative manner. 
With prior knowledge of phonetics, IPA-CLIP learns both phoneme embeddings and pronunciation embeddings through distillation.

In these experiments, we compare the performance of IPA-CLIP with a baseline method. The baseline employs a pronunciation encoder that utilizes an ordinary character-level (phoneme-level) embedding layer instead of the IPA-based one. 
Thus, its obtained embedding space does not consider the phonetic relationships on the IPA chart and implicitly learns such relationships only from the co-occurrences of phonemes in the training data.
In addition, we also explore whether the weights of the feature matrix $W$ of IPA-CLIP should be trainable or frozen as randomly initialized values.
If frozen, the mapping that $W$ is responsible for becomes identical to a random mapping. If trainable, it reflects the phonetic relationships learned from co-occurrences of phonemes, which could be both a boon and a bane. 
In the following sections, we distinguish these two cases by referring to them as \textbf{\textit{Proposed (Trainable)}} and \textbf{\textit{Proposed (Frozen)}}.
In both cases, the weights of the DistilBERT and the additional linear layer are always trained.

Note that the experiments described in the following sections measure performance only towards phonemes that appear in English to perform a fair comparison of the proposed IPA-CLIP with the baseline.

\subsection{Experiment on Phoneme Spaces}\label{sec:exp1:metric}

The experiment in this section compares the proposed IPA-based embedding (proposed methods) with the conventional phoneme-level embedding (baseline method). We measure the following characteristics of the learned phoneme embedding spaces with different metrics: (1)~How distinct the distributions of the consonant cluster and the vowel cluster are, (2)~How the consonant layout represents the phonetic relationships among consonants, and (3)~How the vowel layout represents the phonetic relationships among vowels.

The clear distinction between consonants and vowels makes IPA-CLIP easier to distinguish the two types of phonemes. Also, the accordance of the phoneme layouts with the IPA Chart means that IPA-CLIP can calculate the distance among phonemes based on phonetic similarity. We use different metrics between consonants and vowels because, as explained in Section~\ref{sec:method:ipa}, the consonant attributes are categorical while some of the vowel attributes are continuous.

\begin{table*}[t]
\caption{Quantitative evaluation of the phoneme embedding spaces. The silhouette coefficient (Silhouette), the mean Average Precision (mAP), and Spearman's rank correlation (Rank Corr.) denote the distinctness between consonants and vowels, consistency of consonant distributions with phonetics, and correlation of vowel distributions to phonetics, respectively.
}

\label{tab:result}
\begin{center}
\scalebox{0.9}{
\begin{tabular}{clwr{1cm}wr{1cm}wr{1cm}wr{1cm}wr{1cm}wr{1cm}wr{1cm}wr{1cm}r}
\toprule
& & \multicolumn{2}{c}{Silhouette $\uparrow$} & \multicolumn{3}{c}{mAP $\uparrow$ (Consonant)} & \multicolumn{3}{c}{Rank Corr. $\uparrow$ (Vowel)} & \multicolumn{1}{c}{Loss $\downarrow$} \\ \cmidrule(l){3-4} \cmidrule(l){5-7} \cmidrule(l){8-10} \cmidrule(l){11-11}
Base & \multicolumn{1}{c}{Method} & \multicolumn{1}{c}{$s_{C_{c}}$} & \multicolumn{1}{c}{$s_{C_{v}}$} & \multicolumn{1}{c}{Voicing} & \multicolumn{1}{c}{Place} & \multicolumn{1}{c}{Manner} & \multicolumn{1}{c}{Height} & \multicolumn{1}{c}{Back} & \multicolumn{1}{c}{Round} & \multicolumn{1}{c}{MSE} \\ \midrule
\multicolumn{1}{c}{\multirow{3}{*}{\rotatebox[origin=c]{90}{ViT-B32}}}
& Baseline & $-0.006$ & 0.049 & 0.585 & 0.433 & 0.394 & 0.524 & 0.574 & 0.796 & \textbf{0.0084} \\
& Proposed (Trainable) & 0.036 & 0.193 & \textbf{0.753} & 0.763 & 0.717 & \textbf{0.913} & 0.666 & 0.882 & 0.0096 \\
& Proposed (Frozen) & \textbf{0.252} & \textbf{0.558} & 0.735 & \textbf{0.812} & \textbf{0.845} & 0.889 & \textbf{0.688} & \textbf{0.925} & 0.0092 \\ \midrule
\multicolumn{1}{c}{\multirow{3}{*}{\rotatebox[origin=c]{90}{ViT-L14}}}
& Baseline & $-0.014$ & 0.054 & 0.589 & 0.421 & 0.342 & 0.541 & 0.592 & 0.753 & \textbf{0.027} \\
& Proposed (Trainable) & $-0.036$ & 0.217 & 0.705 & 0.767 & 0.642 & \textbf{0.891} & 0.680 & \textbf{0.925} & 0.028 \\
& Proposed (Frozen) & \textbf{0.233} & \textbf{0.568} & \textbf{0.735} & \textbf{0.810} & \textbf{0.837} & 0.890 & \textbf{0.685} & \textbf{0.925} & 0.028 \\ \midrule
& Upper Bound & 1.000 & 1.000 & 1.000 & 1.000 & 1.000 & 1.000 & 1.000 & 1.000 & --- \\ \bottomrule
\end{tabular}
}
\end{center}
\end{table*}

\subsubsection{Distinctness of Consonant and Vowel Distributions}

To measure the distinctness of consonants and vowels on the phoneme embedding space, we calculate the silhouette coefficient~\cite{bib:silhouette} between the consonant and vowel clusters on the embedding space. 
Given that $C_{c}$ (respectively $C_{v}$) is a set of consonants (vowels), $c$ is an element of $C_{c}$,  and $x_{c}$ is the embedding vector of $c$,
the coefficient $s_{c}$ for the consonant $c$ is calculated as 
\begin{equation}
    s_{c} = \frac{b_{c}-a_{c}}{\max(a_{c}, b_{c})},
\end{equation}
where 
\begin{equation}
a_{c} = \frac{1}{|C_{c}|-1} \sum_{\hat{c} \in C_{c},c \neq \hat{c}} d(x_{c}, x_{\hat{c}}), \,
b_{c} = \frac{1}{|C_{v}|} \sum_{v \in C_{v}} d(x_{c}, x_{v}),
\end{equation}
and $d$ is the Euclidean distance. 
The silhouette coefficient for the consonant cluster, $s_{C_{c}}$, is then calculated by averaging the coefficients among all consonants.
The coefficient for the vowel cluster, $s_{C_{v}}$, is also achieved in the same way, by swapping $c$ and $v$ in the equations.

$s_{C_{c}}$ (respectively $s_{C_{v}}$) ranges between $[-1,1]$, where a high value indicates that the consonant (vowel) cluster is well distinct from the vowel (consonant) cluster.

\subsubsection{Consistency of Consonant Distribution with Phonetics}

To measure the consistency of the consonant distribution on the phonetic space and the consonant categorization on the IPA chart, we calculate the mean Average Precision (mAP) for each consonant attribute. It is a mean of Average Precision scores (AP), which is formalized using Precision@$k$ and Recall@$k$ scores as 
\begin{equation}
    \mathrm{AP} = \sum_{k=1}^{|C_{c}|} \mathrm{Precision}@k \{\mathrm{Recall}@(k)-\mathrm{Recall}@(k-1)\}.
\end{equation}
We then rank the retrieved consonants by measuring the Euclidean distance on the phonetic space. 
Throughout the calculation, we regard consonants that share the focused attribute as \textit{relevant}. For instance, when the voiced bilabial plosive \textipa{/b/} is evaluated in terms of \textit{voicing} attribute, the set of its \textit{relevant} consonants, $R$, then becomes a set of all voiced consonants containing e.g. \textipa{/d/}, \textipa{/m/}, and \textipa{/g/}.
If the retrieved ranking for the consonant \textipa{/b/} is [\textipa{/p/},\textipa{/d/},\textipa{/t/},\textipa{/m/},\textipa{/g/}, \dots] in order, the $\mathrm{AP}$ score for \textipa{/b/}, $\mathrm{AP}_{\textrm{\textipa{b}}}$, will be 
\begin{equation}
\begin{aligned}
    \mathrm{AP}_{\textrm{\textipa{b}}} &= \frac{0}{1} \left( \frac{0}{|R|}-\frac{0}{|R|}\right) + \frac{1}{2}\left (\frac{1}{|R|}-\frac{0}{|R|}\right) + \frac{1}{3}\left (\frac{1}{|R|}-\frac{1}{|R|}\right) \\ 
    &+ \frac{2}{4}\left (\frac{2}{|R|}-\frac{1}{|R|}\right) + \frac{3}{5} \left (\frac{3}{|R|}-\frac{2}{|R|}\right)+\cdots \\
    &= \frac{1}{|R|}\left (\frac{1}{2}+\frac{2}{4}+\frac{3}{5}+\cdots\right).
\end{aligned}
\end{equation}
Lastly, the mAP metric for each consonant attribute is calculated by averaging the AP scores among all consonants.
Thus, a high mAP metric means that phonetically similar phonemes in terms of the target attribute are located close to each other on the phoneme space.

\subsubsection{Correlation of Vowel Distribution to Phonetics}

\begin{figure*}[t]
  \begin{tabular}{cc}
    \begin{minipage}[t]{0.55\hsize}
    \begin{center}
      \includegraphics[width = 1\columnwidth]{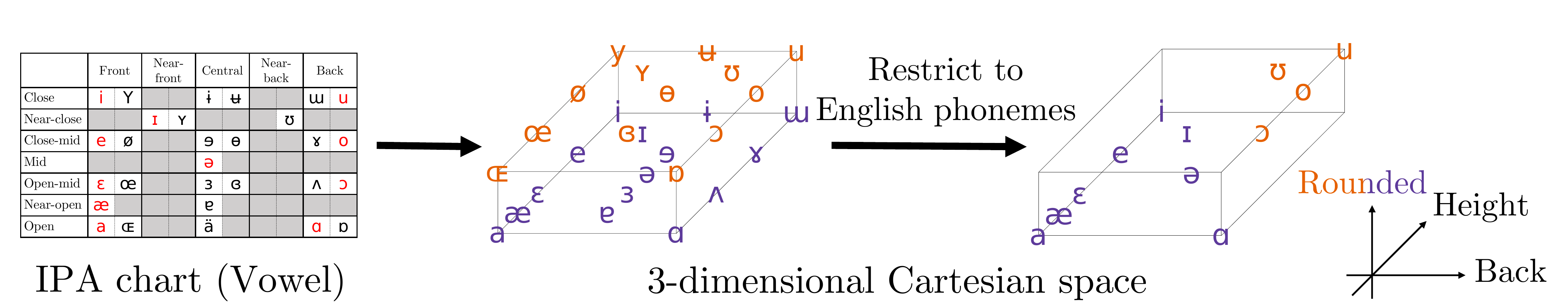}
      \subcaption{Conversion of the IPA chart to 3-dimensional Cartesian space.}
      \label{fig:rcorr:a}
    \end{center}
    \end{minipage}
    \begin{minipage}[t]{0.45\hsize}
    \begin{center}
      \includegraphics[width = 1\columnwidth]{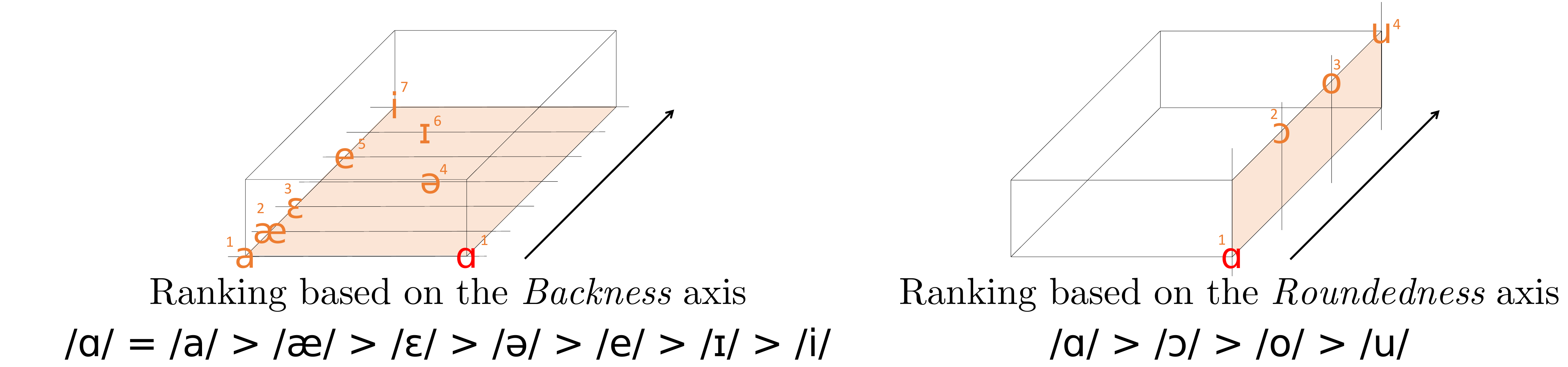}
      \subcaption{Calculation of ground truth rankings of \textipa{/A/} on the \textit{height} axis.}
      \label{fig:rcorr:b}
    \end{center}
    \end{minipage}
  \end{tabular}
  \caption[]{Core ideas of (a) converting the IPA chart to a 3-dimensional Cartesian space and (b) creating ground truth vowel rankings to measure rank correlation scores.}
  \label{fig:rcorr}
\end{figure*}    

This section describes how we measure the correlation between the vowel distribution on the phonetic space and the vowel order on the IPA chart.

The core idea is illustrated in Fig.~\ref{fig:rcorr}. First, inspired by Bay et al.~\cite{bib:bay}, we map every vowel onto the three-dimensional Cartesian space that replicates the IPA chart. The three axes of the Cartesian space represent vowel attributes of \textit{height}, \textit{backness}, and \textit{roundedness}, respectively. Next, for each of the three axes/attributes, we calculate Spearman's rank correlation between the vowel distribution on this Cartesian space and that on the phonetic space. 
We create two ground truth rankings for each attribute by sorting vowels that share one of the other two attributes. For instance, as illustrated in Fig.~\ref{fig:rcorr}, when evaluating the vowel \textipa{/A/} on the \textit{height} attribute, we calculate the following two rankings: (1)~the ranking among the back vowels: \textipa{/A/} $>$ \textipa{/O/} $>$ \textipa{/o/} $>$ \textipa{/u/}, and (2)~the ranking among the unrounded vowels: \textipa{/A/} $=$ \textipa{/a/} $>$ \textipa{/\ae/} $>$ \textipa{/E/} $>$ \textipa{/@/} $>$ \textipa{/e/} $>$ \textipa{/I/} $>$ \textipa{/i/}.

With these ground truth rankings prepared, we then calculate two rank correlation scores between each of the two rankings of the target vowel and the ranking of the similar vowels retrieved on the phonetic space.
The retrieval is performed on the basis of the Euclidean distance.
Lastly, for each vowel attribute, we compute the average value among all vowels to be the rank correlation metric on the phonetic space.
Thus, a high rank correlation metric means that the distribution of vowels on the phoneme space represents the phonetic relationships among vowels well.

\subsubsection{Results and Discussion}
The results of this experiment are shown in Table~\ref{tab:result}. The table also showcases the validation loss at the point of 50 epochs.
Overall, \textit{Proposed (Frozen)} performs best in almost all metrics. \textit{Proposed (Trainable)} is also comparable except for the silhouette coefficient. The baseline method performs worse than the proposed methods with the exclusion of the slightly better validation loss.

The great advantage of \textit{Proposed (Frozen)} to the other methods is the silhouette coefficient. 
A high silhouette coefficient means that the distributions of the consonant and the vowel clusters are distinct and thus have little overlap on the phoneme embedding space.
Since the coefficient drops through the training of the weight matrix in the baseline and the \textit{Proposed (Trainable)} methods, this comparison points out that the embeddings learned from the phoneme co-occurrences in sentences do not clearly distinguish consonants and vowels. This contradicts the fact that such neural embeddings are known to represent phonetic relationships quite well~\cite{bib:kolachinamagyar,bib:boldsen}.
Yet, since even the baseline performs moderately in the mAP and rank correlation metrics, we can also confirm that the neural embeddings can learn relationships within consonants and within vowels even without any explicit prior.

Moreover, between the baseline and the two proposed methods, the increase of both mAP and rank correlation is confirmed. This suggests that the proposed IPA-based embedding works quite well in differentiating both within consonants and within vowels. In contrast, the slight deceleration of the decrease of the validation loss indicates that this inserted prior functions as an additional restriction for the model, impairing the optimization during the distillation process.

\begin{figure}[t]
  \begin{tabular}{cc}
    \begin{minipage}[t]{0.48\hsize}
      \centering
      \includegraphics[width = 1\columnwidth]{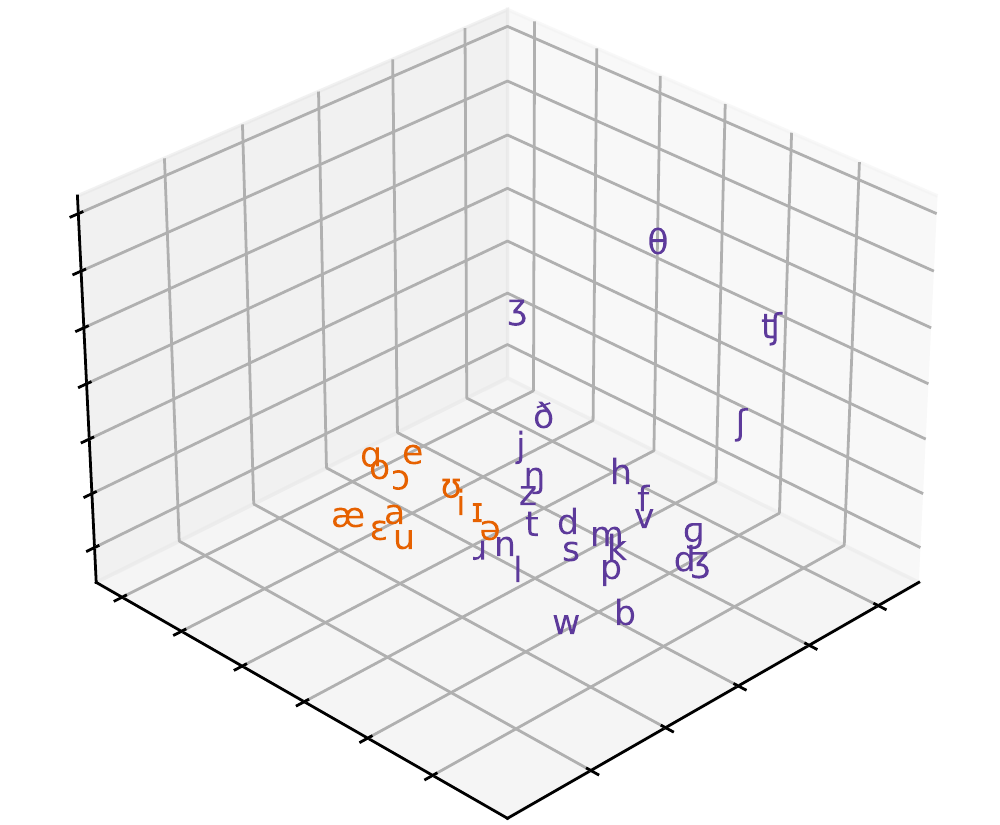}
      \subcaption{Baseline}
      \label{fig:visualization:baseline}
    \end{minipage} &
    \begin{minipage}[t]{0.48\hsize}
      \centering
      \includegraphics[width = 1\columnwidth]{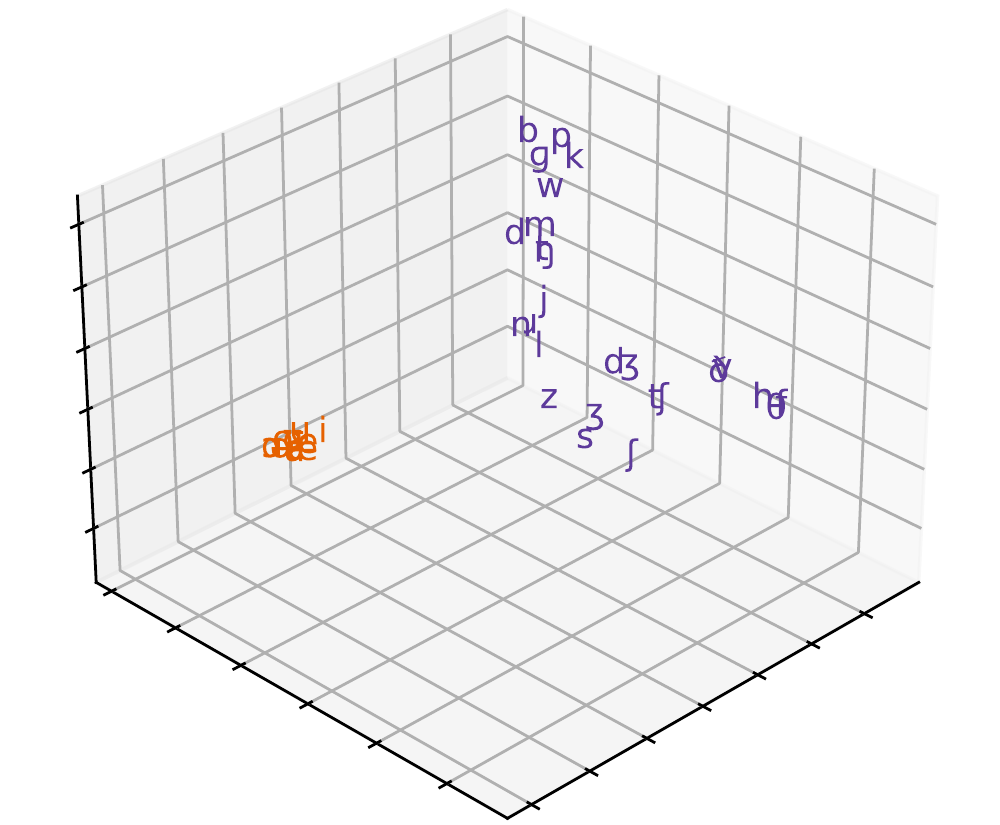}
      \subcaption{Proposed (Frozen)}
      \label{fig:visualization:frozen}
    \end{minipage}
  \end{tabular}
  \caption[]{Visualization of the phoneme embedding spaces. Consonants and vowels are shown in blue and yellow, respectively, to compare their distributions.}
  \label{fig:visualization}
\end{figure}

The comparison with the upper bounds spotlights that the silhouette coefficients for the consonant cluster $s_{c}$, even of the \textit{Proposed (Frozen)} method, are not very high. One possible reason for this is that the consonants distribute more widely than vowels regardless of the degree of the distinctness of the two clusters. To verify this, we visualized the phoneme spaces of the baseline and the \textit{Proposed (Frozen)} methods.  Figure~\ref{fig:visualization} shows a scatter plot of all consonants and vowels on the three-dimensional spaces suppressed by Principal Component Analysis (PCA). As expected, in both methods, the consonant cluster distributes more widely than the vowel cluster, which could have vastly affected the silhouette coefficient metrics. Nevertheless, the figure still reveals that the space of the proposed method is more reasonably ordered than that of the baseline method in terms of both the separation of the two clusters and their accordance with the IPA chart.

\subsection{Experiments on Pronunciation Spaces}\label{sec:exp2:metric}

We conduct further experiments to discuss the applicability of IPA-CLIP in multimodal tasks. We measure its performance in the following three retrieval tasks: 
(1)~Retrieval-based image classification from the pronunciations of existing words, 
(2)~Image retrieval from the pronunciations of nonwords, and 
(3)~Text retrieval from the pronunciations of nonwords.
Note that the nonwords here denote such words that do not exist in English but sound similar to certain existing words.

By measuring the performance on these three tasks, we can evaluate the accordance of the pronunciation encoder 
(1)~with the image encoder for existing words, 
(2)~with the image encoder for nonwords, and 
(3)~with the text encoder for nonwords, respectively.

In all of these experiments, we utilize the ImageNet~\cite{bib:imagenet} validation dataset as a source of image-text pairs. The dataset provides 50 images for each of the 1,000 pre-defined classes. We convert each class label into its pronunciation in the same way as we did in Section~\ref{sec:method:implement}. By removing the classes where we failed in converting the labels to their pronunciation, we obtain 912 classes attached with 50 images each in total. Note that we use the class labels identical to the ones used by the authors of CLIP~\cite{bib:clip}, which differ from the class labels that ImageNet provides. The authors modified the labels to reduce the ambiguity of the original ImageNet labels for their retrieval-based image classification task.

Besides, the following experiments only use the models trained upon CLIP ViT-L14 since the experiment in Section~\ref{sec:exp1:metric} showed no significant difference among the choice of the base models. 
Also, we do not compare the performance of the proposed methods with models other than CLIP~\cite{bib:clip}. This is because all of the methods introduced in Section~\ref{sec:related:computational} are not applicable to multimodal retrieval tasks, and the other CLIP-based models introduced in Section~\ref{sec:related:clipextension}~\cite{bib:multilingualclip, bib:wav2clip} were outperformed by CLIP in these tasks. 

\subsubsection{Image Classification from Pronunciation}

We experiment on a retrieval-based image classification task similar to the one in a previous study~\cite{bib:clip}. 
Here, IPA-CLIP classifies an image by measuring the cosine similarities between the embedding of the image and the embeddings of the class labels in the form of, e.g., \textipa{/@ "foU""toU @v} <CLASS>\textipa{/} (``A photo of <CLASS>''). For example, given an image and two class labels ``Dog'' and ``Cat'', IPA-CLIP first calculates the embedding of the image and the embeddings of the two pronunciations \textipa{/@ "foU""toU @v dOg/} and \textipa{/@ "foU""toU @v k\ae t/}. It then measures the cosine similarity between the image embedding and each of the pronunciation embeddings.
By finding the image-label pair that gives the maximum similarity, it determines the class label of the image.

In this experiment, we also filter out classes by measuring the word frequency (as the Zipf scale, which we call \textit{Zipf frequency}) of their labels using an existing Python package~\cite{bib:wordfreq}. The aim of this is to observe how the rare class labels, which would never appear or appear few in the distillation process of IPA-CLIP, affect the classification results.

We measure accuracy scores as a metric. We compare the performance of our methods with CLIP~\cite{bib:clip}, which classifies images from text labels using the text encoder. To see the difference in the functionality of IPA-CLIP pronunciation encoder and the CLIP text encoder, we also attempt to merge the two by taking the average of their embeddings on the joint embedding space. We call this ``\textbf{\textit{Proposed (Frozen) + CLIP}}''. 

\subsubsection{Nonword-to-Image Retrieval}\label{sec:exp2:pronuntoimage}

To evaluate the robustness of IPA-CLIP towards nonsense words having certain similar-sounding existing words, we prepare a set of nonwords by slightly modifying the class labels of ImageNet. 

First, we focus only on the labels whose Zipf frequency is three or more (297 classes).
Then, for labels starting with a sole consonant (216 classes satisfy this), we substitute the initial consonant with other possible consonants (e.g., from \textipa{/dEsk/}: ``Desk'' to \textipa{/zEsk/}, \textipa{/nEsk/}, etc.) to make a set of nonwords which sound similar to the original word.
Next, we remove the generated words that happen to exist in the English vocabulary. To check this, we use the SCOWL wordlist and the CMU dictionary\footnote{\url{https://github.com/menelik3/cmudict-ipa/} (Accessed Jan. 19, 2023)}.
This process yields 3,530 nonwords stemming from either of the 216 classes. 

During this preparation, we also prepare the text equivalents by automatically converting each phoneme into its spelling (``Zesk'' for \textipa{/zEsk/}) so that we can also evaluate the text-based original CLIP.
Table~\ref{tab:result:consonants} lists all consonants used for the substitution along with their spelling correspondents.
As shown in the table, the candidate consonants are selected from all consonants appearing at the beginning of English words, except for \textipa{/D/}, which becomes identical to \textipa{/T/} when spelled.

Using these nonwords, we perform a nonword-to-image retrieval task. Given a nonword, the objective is to retrieve the images belonging to the class from which the nonword stems. For instance, given the nonword \textipa{/zEsk/}, we measure how many of the 50 images in the class ``Desk'' can be retrieved from the pronunciation embedding of \textipa{/@ "foU""toU @v zEsk/}.

We measure Recall@50 scores as a metric. We split the evaluation based on how phonetically similar the nonword is to its original word. Regarding this, we observe the number of shared attributes between the first consonant in the nonword and that of its original word. The aim of this is to assess whether the methods capture the phonetic similarity among consonants. Giving always similar scores regardless of the number of shared attributes mean that the method does not consider phonetic similarity. In contrast, if the score correlates to the number of shared attributes, it can be said that the method associates nonwords with their similar-sounding words based on phonetic similarity.

\begin{table}[t]
\caption{Candidate consonants and corresponding spellings used to generate nonwords from the class labels of ImageNet.}
\label{tab:result:consonants}
\begin{center}
\scalebox{0.9}{
\begin{tabular}{l|wc{2.5mm}wc{2mm}wc{2mm}wc{2mm}wc{2mm}wc{2mm}wc{2mm}wc{2mm}wc{2mm}wc{2mm}wc{2mm}}
\toprule
Consonant & \textipa{/s/} & \textipa{/n/} & \textipa{/f/} & \textipa{/l/} & \textipa{/z/} & \textipa{/b/} & \textipa{/\*r/} & \textipa{/p/} & \textipa{/g/} & \multicolumn{2}{c}{\textipa{/k/}} \\
Spelling & `s' & `n' & `f' & `l' & `z' & `b' & `r' & `p' & `g' & \multicolumn{2}{c}{`k' or `c'} \\ \midrule
Consonant & \textipa{/d/} & \textipa{/m/} & \textipa{/T/} & \textipa{/t/} & \textipa{/\textdyoghlig/} & \textipa{/j/} & \textipa{/h/} & \textipa{/v/} & \textipa{/S/} & \textipa{/\textteshlig/} & \textipa{/w/} \\
Spelling & `d' & `m' & `th' & `t' & `j' & `y' & `h' & `v' & `sh' & `ch' & `w' \\ \bottomrule
\end{tabular}
}
\end{center}
\end{table}


\begin{table}[t]
\caption{Accuracies of the image classification on 1,000-class ImageNet~\cite{bib:imagenet} dataset. We only use classes that can be converted to their pronunciation and use Zipf frequency to filter out the classes having less frequent and rare label names.}
\label{tab:result:image-pronun}
\begin{center}
\scalebox{0.9}{
\begin{tabular}{lwr{6mm}wr{6mm}wr{6mm}wr{6mm}}
\toprule
\multicolumn{1}{c}{Zipf Frequency} & $\geq$ 0.0 & $\geq$ 1.5 & $\geq$ 3.0 & $\geq$ 4.5 \\ \midrule
\multicolumn{1}{c}{Number of Classes} & 912 & 492 & 297 & 29 \\ \midrule
Baseline & 0.600 & 0.696 & 0.777 & 0.877 \\
Proposed (Trainable) & 0.581 & 0.686 & 0.769 & 0.886 \\ 
Proposed (Frozen) & 0.590 & 0.683 & 0.764 & 0.885 \\ \midrule
CLIP~\cite{bib:clip} & \textbf{0.712} & 0.751 & 0.765 & 0.891 \\ \midrule
Proposed (Frozen) + CLIP~\cite{bib:clip} & 0.705 & \textbf{0.765} & \textbf{0.799} & \textbf{ 0.897} \\ \bottomrule
\end{tabular}
}
\end{center}
\end{table}

\begin{figure}[t]
  \begin{tabular}{cc}
    \begin{minipage}[t]{0.46\hsize}
      \centering
      \includegraphics[width = 1\columnwidth]{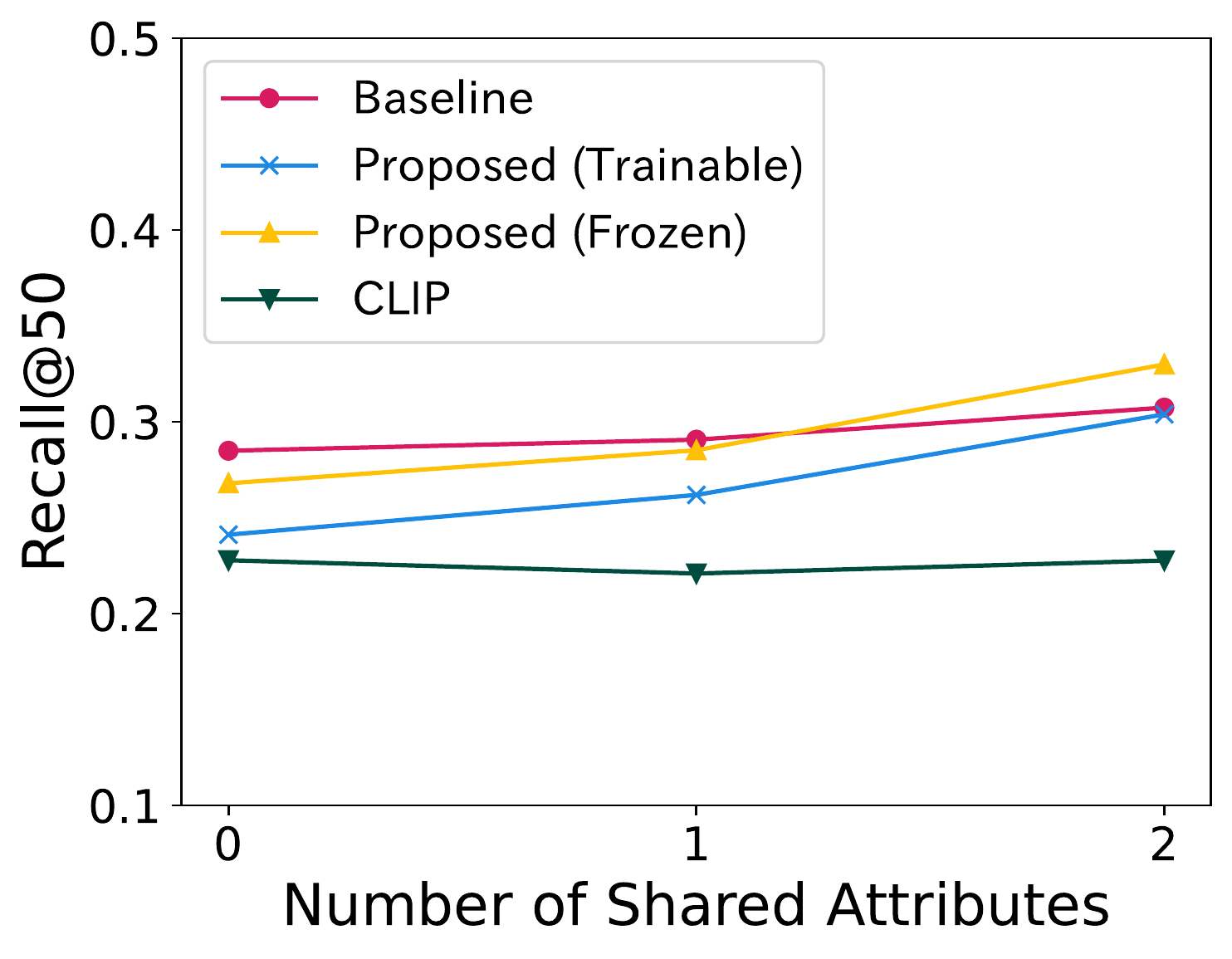}
      \subcaption{Nonword-to-Image}
    \end{minipage} &
    \begin{minipage}[t]{0.46\hsize}
      \centering
      \includegraphics[width = 1\columnwidth]{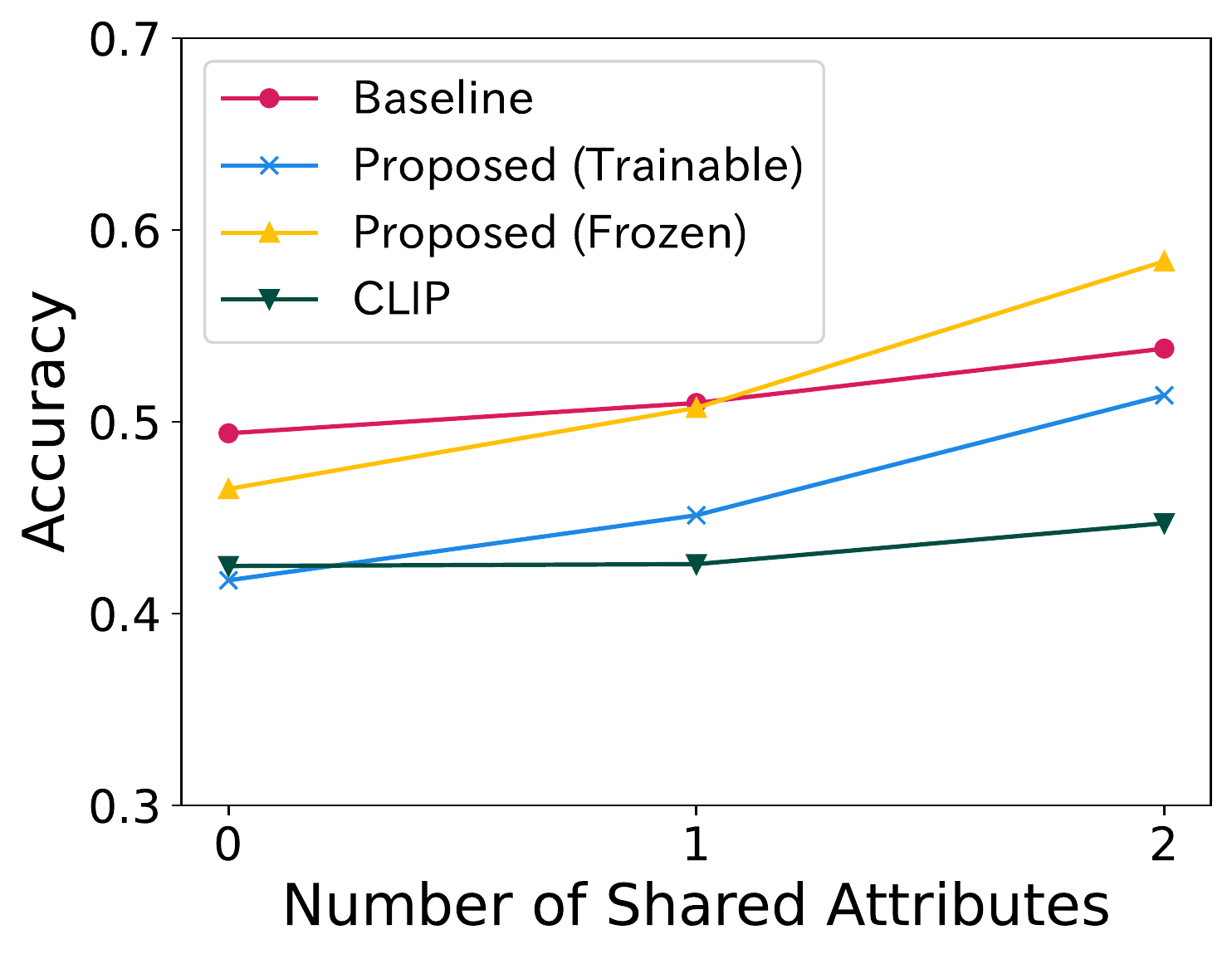}
      \subcaption{Nonword-to-Text}
    \end{minipage}
  \end{tabular}
  \caption[]{Results of (a) image retrieval and (b) text retrieval from nonwords written with either phonetic symbols (Baseline and Proposed) or Latin alphabet letters (CLIP). High accordance with phonetic similarity yields a high correlation between the number of shared attributes (x-axis) and both metrics (y-axis).}
  \label{fig:nonword-to}
\end{figure}

\subsubsection{Nonword-to-Text Retrieval}

The procedure of the nonword-to-text retrieval is quite similar to the one described in Section~\ref{sec:exp2:pronuntoimage}, but the retrieval in this experiment targets texts instead of images.
We use the set of 3,530 nonwords for 216 classes prepared in Section~\ref{sec:exp2:pronuntoimage}. In this experiment, the models retrieve the text of the class from which the nonword stems. For example, given the nonword \textipa{/zEsk/}, we evaluate whether the methods can retrieve the text embedding of the sentence ``A photo of desk'' among the text embeddings of 216 classes.
We measure accuracy scores as a metric.

\subsubsection{Results and Discussion}

First, the results of the image classification are shown in Table~\ref{tab:result:image-pronun}. 
The table shows that the performance of the proposed methods is vastly affected by the frequency/rareness of the class labels. As can be seen on the left side of the table, they perform much worse than CLIP~\cite{bib:clip} when the classes contain rare words. This is mainly because these models, as student models, have not been exposed much to these rare words during the distillation process. In contrast, as the rare words drop out from the evaluation, their performance comes to be comparable to CLIP. 

The most interesting is ``\textit{Proposed (Frozen) + CLIP}''. Despite its simple fusion strategy of the two modalities, it performs the best in almost all settings. This indicates the effectiveness of introducing the pronunciation modality into existing V\&L pretrained models. Looking into the 297-class confusion matrix revealed the strengths and weaknesses of each encoder, although we do not list the actual matrix to conserve space.
The pronunciation encoder is more sensitive to pronunciation differences, while the text encoder is stronger against the meaning gaps. 
For example, \textit{Proposed (Frozen)} misclassified ``\textit{Block plane}'' as ``\textit{Buckle}'' since they sound similar. In contrast, CLIP misclassified ``\textit{Slug}'' as ``\textit{Snail}'' since their meanings are quite similar. Since ``\textit{Proposed (Frozen) + CLIP}'' correctly classified both cases, averaging the embeddings of the two encoders could have compensated for their weaknesses.

Next, Fig.~\ref{fig:nonword-to} shows the results of the nonword-to-image and nonword-to-text retrieval tasks, respectively.
The tendency of the results is similar throughout the two separate tasks: The baseline method retrieves the best when the number of shared attributes is 0 or 1, while \textit{Proposed (Frozen)} performs best when it is 2. 
This suggests that \textit{Proposed (Frozen)} is likely to associate nonwords with the original word only when the words are phonetically similar.
Therefore, this confirms that \textit{Proposed (Frozen)} considers the phonetic similarity between consonants more accurately than the other methods also in the pronunciation embedding space.

We also observed that the proposed methods always perform better than CLIP in these nonword-centered tasks. \textit{Proposed (Frozen)} improves Recall@50 by 0.112 points and accuracy by 0.137 points when the number of shared attributes is 2. This fact verifies that the proposed pronunciation modality makes CLIP more robust against nonwords.

\begin{figure}[t]
  \begin{center}
      \includegraphics[width = 0.95\columnwidth]{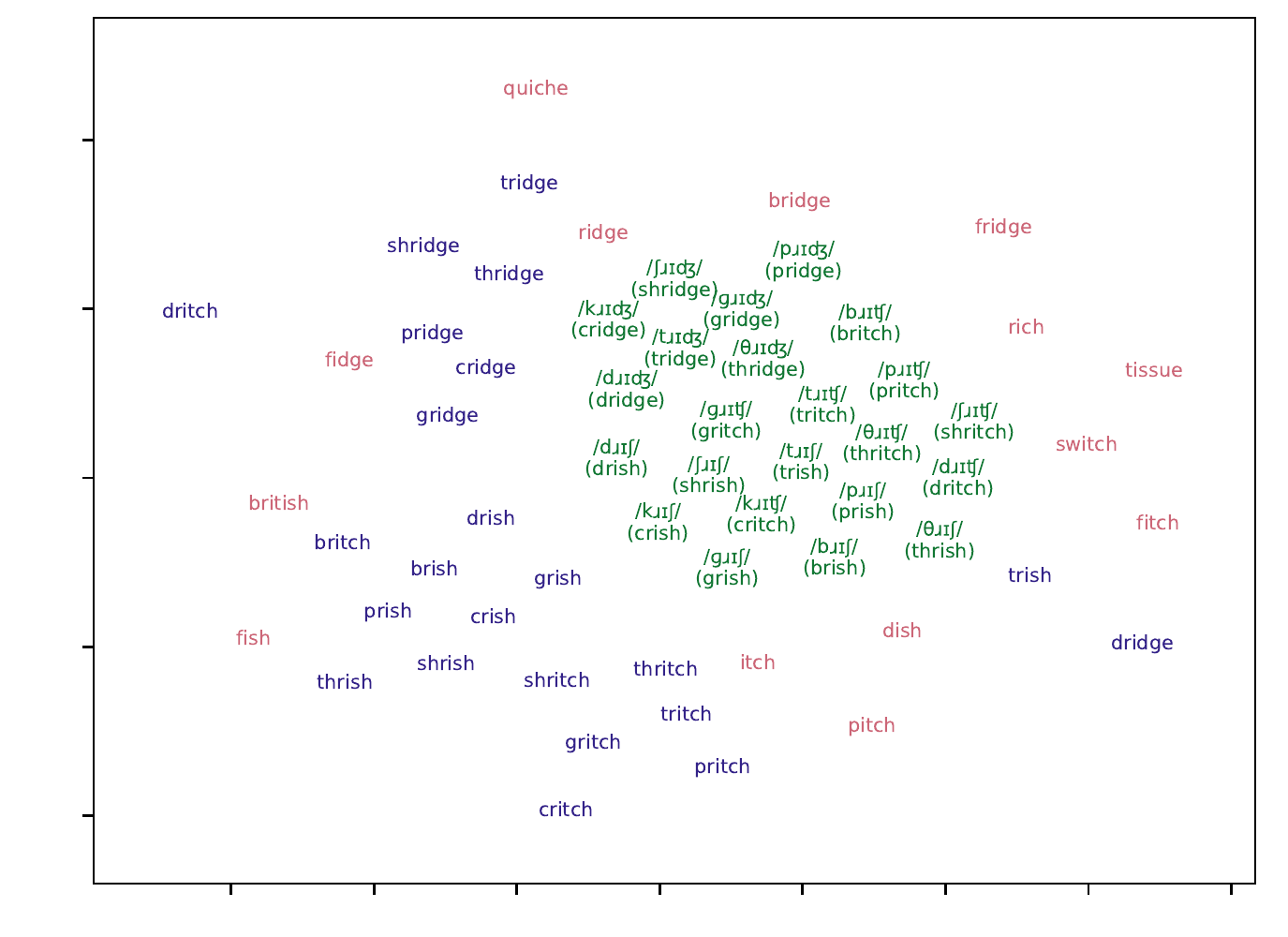}
  \end{center}
  \caption[]{t-SNE~\cite{bib:tsne} visualization of the text and pronunciation embeddings of nonwords. Text embeddings of existing words (red) and nonwords (purple) are calculated by CLIP, while pronunciation embeddings of nonwords (green) are calculated by IPA-CLIP.}
  \label{fig:interpolate}
\end{figure}

To further analyze the difference between CLIP and IPA-CLIP, we visualized how words and nonwords are distributed on their shared embedding space. The scatter plot, shown in Fig.~\ref{fig:interpolate}, illustrates the embeddings of existing words and nonwords sounding similar to ``\textit{Bridge}'', calculated by either CLIP or IPA-CLIP.
It reveals that IPA-CLIP places nonwords such as ``\textit{Pridge}'' (\textipa{/p\*rI\textdyoghlig/}) and ``\textit{Britch}'' (\textipa{/b\*rI\textteshlig/}) in positions very close to their similar-sounding existing word ``\textit{Bridge}'' (\textipa{/b\*rI\textdyoghlig/}). 
In contrast, CLIP does not place any nonwords, even ``\textit{Pridge}'', near ``\textit{Bridge}''. 
This supports the results that IPA-CLIP considers the phonetic similarity of words. 

\section{Qualitative Evaluation}\label{sec:exp2}

We also evaluate how much the proposed methods attune the CLIP embedding space to actual human perception regarding pronunciation similarity.
This evaluation uses the pronunciation similarity rankings collected by Vitz and Winkler~\cite{bib:vitzwinkler}.
In each of their four trials, native American English speakers rated the pure sound similarity between a given target word and each of its 25 comparison words.
Their four experiments differ only in the target word, which is ``\textit{Sit}'', ``\textit{Plant}'', ``\textit{Wonder}'', and ``\textit{Relation}'', respectively, as well as its comparison words.
In the first three experiments, comparison words are a mixture of valid and nonsense English words that have a similar syllable structure as the target word.
In the last experiment, this constraint for the syllable structure is removed. 
For example, the comparison words for the target word ``\textit{Sit}'' include ``\textit{Pit}'', ``\textit{Sat}'', and ``\textit{Tass}'', and those for the target word ``\textit{Relation}'' include ``\textit{Belation}'', ``\textit{Fascinating}'', and ``\textit{Get}''.

In our evaluation, given a target word, we first calculate the cosine similarity between the word and each of its comparison words on the pronunciation space of each method to create a similarity ranking. Then, we measure its rank correlation to the ground truth as a metric.
A higher value means that the embedding space better fits human perception regarding pronunciation similarity.

The results are shown in Table~\ref{tab:result2}.
All of the pronunciation-based methods outperform the conventional text-based CLIP, which verifies that the phonetic prior forces similar-sounding words to become closer to each other. 
Within the pronunciation-based methods, the performance of \textit{Proposed (Frozen)} is particularly bad when the target word is ``\textit{Sit}''. This is due to the short syllable length of its comparison words, yielding much more possible similar-sounding words than the other target words.
Thus, its phonetic knowledge could have disturbed the calculation of the embeddings of such short syllable words, which would be a shortcoming of the proposed approach.
Nevertheless, since this evaluation covers just these four specific cases, the results do not spotlight which of the pronunciation-based methods works best in general.

\begin{table}[t]
\caption{Qualitative evaluation of the pronunciation embedding space. The scores denote rank correlations between the word similarity measured on the embedding space of each method and the ground truth rated by humans.}
\label{tab:result2}
\begin{center}
\scalebox{0.9}{
\begin{tabular}{lwr{12mm}wr{12mm}wr{12mm}wr{12mm}}
\toprule
\multicolumn{1}{c}{Target Word} & \multicolumn{1}{c}{Sit} & \multicolumn{1}{c}{Plant} & \multicolumn{1}{c}{Wonder} & \multicolumn{1}{c}{Relation}\\ \midrule
Baseline & 0.535 & 0.397 & \textbf{0.693} & 0.442 \\
Proposed (Trainable) & \textbf{0.642} & \textbf{0.549} & 0.526 & 0.485 \\
Proposed (Frozen) & 0.385 & 0.420 & 0.640 & \textbf{0.504} \\ \midrule
CLIP~\cite{bib:clip} & 0.353 & 0.402 & 0.585 & 0.304 \\ \bottomrule
\end{tabular}
}
\end{center}
\end{table}

\section{Conclusion}\label{sec:conclusion}

We proposed an IPA-based phoneme embedding, which integrates the phonetic relationships on the IPA (International Phonetic Alphabet) chart into a character/phoneme-level embedding. We also proposed IPA-CLIP, which extends the Vision and Language (V\&L) pretrained model CLIP to accept an arbitrary pronunciation input by adopting the IPA-based phoneme embedding. This enables it to process inputs even if they contain nonsense words (nonwords).  
The proposed IPA-CLIP model was trained by distilling the text encoders of CLIP ViT-B/32 and ViT-L/14 models. Evaluations showed the effectiveness of adopting the IPA-based phoneme embedding against neural embeddings obtained without phonetic priors. We also evaluated the performance of IPA-CLIP on some multimodal retrieval tasks, showing not only the comparable performance of IPA-CLIP but also its potential to outperform the original text-based CLIP. In the setting where nonwords are input, IPA-CLIP performs always better than CLIP, verifying the robustness of IPA-CLIP against nonwords. 
Further evaluation verified the correlation between the pronunciation embedding space of IPA-CLIP and human perception regarding pronunciation similarity.
These results guarantee the usefulness of IPA-CLIP in enhancing language understanding of multimedia systems.

For future work, further analysis is needed to investigate under which conditions the proposed approach has advantages over text-based methods.
Besides, we plan to create a system based on the proposed method that models human perception toward nonwords. Moreover, we are also interested in expanding IPA-CLIP to multiple languages to quantize the inter-lingual difference in human perception. In the face of this, one limitation we recognize is that the proposed IPA-based embedding ignores the auditory feature of pronunciation, such as the auditory similarity between phonemes, tones, and prosody. To tackle this limitation, one possibility is to create a CLIP extension to audio inputs, like the ones already accomplished in previous studies~\cite{bib:audioclip,bib:wav2clip}, and combine it with the proposed IPA-CLIP.

\begin{acks}
This work was partly supported by Microsoft Research CORE16 program and JSPS Grant-in-aid for Scientific Research (22H03612). 
This work is a fruit of a joint research project between Nagoya University and the National Institute of Informatics.

The first author would like to take this opportunity to thank the ``Nagoya University Interdisciplinary Frontier Fellowship'' supported by Nagoya University and JST, the Establishment of University Fellowships towards the Creation of Science Technology Innovation, Grant Number JPMJFS2120.
\end{acks}

\bibliographystyle{ACM-Reference-Format}


\end{document}